# A Clinical and Finite Elements Study of Stress Urinary Incontinence in Women Using Fluid-Structure Interactions

Mojtaba Barzegari[1], Bahman Vahidi[1]*, Mohammad Reza Safarinejad[2]

1- Division of Biomedical Engineering, Department of Life Science Engineering, Faculty of New Sciences and Technologies, University of Tehran, Tehran, Iran
2- Clinical Center for Urological Disease Diagnosis and Private Clinic Specialized in Urological and Andrological Genetics, Tehran, Iran

* P.O.B. 1439957131, Tehran, Iran, bahman.vahidi@ut.ac.ir

**Abstract**

Stress Urinary Incontinence (SUI) or urine leakage from urethra occurs due to an increase in abdominal pressure resulting from stress like a cough or jumping height. SUI is more frequent among post-menopausal women. In the absence of bladder contraction, vesical pressure exceeds from urethral pressure leading to urine leakage. Despite a large number of patients diagnosed with this problem, few studies have investigated its function and mechanics. The main goal of this study is to model bladder and urethra computationally under an external pressure like sneezing. Finite Element Method and Fluid-Structure Interactions are utilized for simulation. Linear mechanical properties assigned to the bladder and urethra and pressure boundary conditions are indispensable in this model. The results show good accordance between the clinical data and predicted values of the computational models, such as the pressure at the center of the bladder. This indicates that numerical methods and simplified physics of biological systems like inferior urinary tract are helpful to achieve the results similar to clinical results, in order to investigate pathological conditions.

**Key words**: Computational Fluid Dynamics, Urinary Tract, Stress Urinary Incontinence, Finite Element Method, Fluid-Structure Interaction

**Introduction**

Recently, diseases associated with urine and genital tract, with the general term urology, are more prevalent among both men and women older than 40 years old. The most critical subset of these problems is urinary incontinence [1]. Stress Urinary Incontinence is very common among women, which occurs due to a mechanical pressure like sneezing or jumping height. In this type of incontinence, any activity that increases abdominal pressure (including laughing, coughing, sneezing, and straining) leads to urine leakage as a result of urethral sphincter weakness. Respecting to the published reports on 2001, costs of urinary incontinence treatments exceeded 16.3 billion US dollar. The reports indicate 75% of total costs were spent on diagnosis and treatment for women [1].

SUI is more prevalent in women [1]. An increase in abdominal pressure in the absence of bladder contraction raises the vesical pressure to a level that exceeds urethral pressure, leading to involuntary loss of urine which mainly characterizes SUI. Abdominal pressure increases due to a mechanical incident like laughing, sneezing, jumping height, or any other tension in the body [1]. This explains why the SUI is considered as a mechanical force.

Although the main reason of SUI remains unknown, a large number of physicians believe that SUI is caused by injuries to the pelvic floor neuro-musculature, which mainly happen among those who have given birth vaginally [1]. While not a life threatening condition, SUI can detrimentally impact the quality of life. Up to now, biomechanical studies over female incontinence are dominated by three theories; including pressure transmission theory proposed by Enhoring [2], the integral theory proposed by Petrus and Ulmestan [3], and the hammock theory proposed by Delancy and Ashton-Miller [4]. Moreover, all these three theories seem paradoxical and within themselves, there is no consistency about involved structures and tissues. On the other hand, analytical studies and organic structures' modeling is a practical way to study biomechanical phenomena.

In the current study, complications of the numerical model could challenge the researcher. In the current paper, the urine flow in the urinary tract is investigated. It is more common to investigate a part of urinary tract since scrutinizing all parts is very complex and time-consuming. Here, the modeling is mainly focused on bladder and urethra. The preferred computational parameter is the fluid (urine) pressure which is compared to clinical data. Pressure is the considered parameter since it is available in urodynamic measurement and it has varied measuring methods.

There are few reports of finite element analysis about SUI. Kim proposed a 2-dimensional model to study urethral closure during stress [5]. Kim proposed an axisymmetric model of the urethra and pelvic floor to find proper dimensions for his model. In Kim's model, a catheter was placed in the urethra. This model was developed to analyze active (muscle contraction) or passive (pressure transmission) contributions to urethral closure. The results showed that an active contraction of sphincter muscle and pelvic floor play an important role in urine continence [5]. The findings showed levator ani's contraction and its connectivity to endo pelvic fascia contribute in urine continence [5]. Zhang et al. developed a rather complex finite element model of the urinary tract and the pelvic floor in athlete women while jumping height to examine feasibility and repeatability of finite element model for urinary incontinence [6]. This model included an ideal modeling of the intestine, vagina, rectum, pelvic diaphragm, urinary diaphragm, bladder and abdominal muscles. Although urine leakage was not clearly defined, it was proved that a finite element model of SUI in women, while they are jumping high, could reproduce the clinical data.

Haridas et al. developed a finite element model to study biomechanical characteristics of the inferior urinary tract and the pelvic floor [7]. Their premier investigations were related to the finite element modeling of the vagina, bladder, uterus, bladder, levator ani, endo pelvic fascia and rectum derived from a specific MRI. A good coherence was observed between the modeling results and the ultrasound images. In the latest studies, Spirka et al. developed a finite element model of stress urinary incontinence [8]. The most critical limitation of this model was employing a quasi-solid model for the fluid. To verify the model, they used limited urodynamic data.

In this study, bladder reaction to an external force (abdominal pressure resulting from a cough) is studied in both physiological and pathological conditions. To achieve this goal, two different computational models are considered to investigate urine dynamics in the bladder. Fluid-Structure Interactions (FSI) method is utilized in the computational models. The required data for the assumption of the simulation are extracted from urodynamic measurements. In comparison to the importance of the considered material properties for tissues, the geometry of urinary tract is less important. So simplified models are considered in this study.

The current study works on extending previous studies to model mechanically the inferior urinary tract, besides validating model through comparing to clinical and urodynamic data. It is mentioned before that the preceding models were focused on one mechanism of urinary continence or they modeled urine leakage. This study is focused on the parameters contributed to leakage and compares physiological and pathological cases. Another feature is employing FSI method to improve the reliability of results.

## Methods

Here, the Stress Urinary Incontinence is studied; collected clinical data are divided into pathological and physiological groups. Deliberate sneezing during urodynamic measurements is a comparison criterion between clinical and analytical data. In this study, a cough and its impact on the inferior urinary tract in both physiological and pathological groups are studied. To achieve this goal, two computational models are developed: one without support structure as a pathological one and the other with a support structure which prevents leakage from the bladder outlet resembling physiological groups. The external force is sneezing which occurs in 200 ms. Both models use linearly elastic material properties. The measured parameter for verification is pressure. Total time of the simulation is 200 ms. The initial external pressure is zero at the initial state and remains zero for 10 ms. Then, the pressure increases to its maximum level (which corresponds to urodynamic measurements) and gradually decreases to the initial level. During the process, the pressure in the center of the bladder changes with time.

To model both pathological and physiological conditions, two different geometries are considered. In the physiological model, pelvic diaphragm closes urethra and prevent urine leakage. In the pathological model, the support structure does not exist, leading to bladder dislocation which prevents urethral closure.

For geometry and boundary conditions, data were collected from the continent case for the physiological condition and the incontinent case for the pathological condition. For geometry construction, bladder capacity is considered and data for both boundary conditions and results validation include abdominal pressure and bladder pressure. Also, the sensitivity of the model to different parameters is considered to verify analytical data.

Inferior urinary tract geometry includes the bladder, urethra and the support structure. In the physiological model, support structure exists, while in the pathological model, it is dislocated and it is upper than the bladder outlet so it has no impact on urine continence. So, it is eliminated from the pathological model. Figure 1 and 2 show pathological and physiological models, respectively.

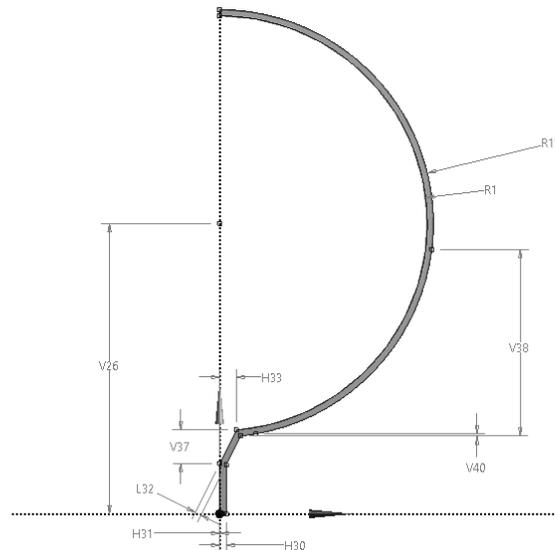

**Fig. 1** Computational geometry for pathological mode

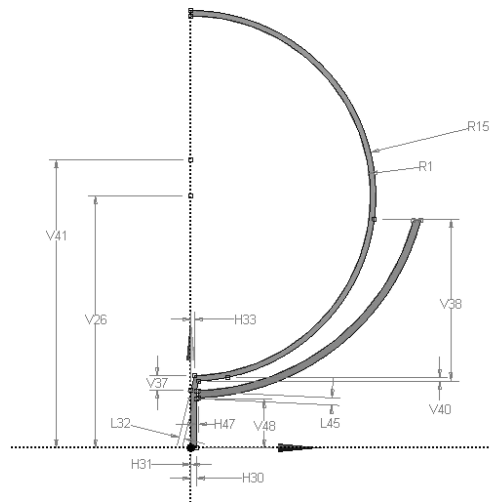

**Fig. 2** Computational geometry for physiological model

Pathological and physiological vesical capacities are 346cc and 410cc, respectively. Since bladder geometry is presumed as a sphere, its diameter changes from 43 mm to 46mm for the pathological and physiological model. Sean et al. studied urodynamic measurement of 42 women and concluded that average thickness of bladder wall with 200cc of capacity would be 1.7 mm [9]. Similar findings presumed that the thickness of bladder wall is approximately 0.9 in 56cc of capacity [5, 10]. With respect to the previous findings [10], to simplify meshing process and fluid and solid net coherence, the bladder wall thickness is presumed 1.5 mm in both conditions.

The urethra is modeled as a cylindrical tube [11, 12]. In both models, the urethra is relatively closed and fluid flow would increase its diameter. Since the purpose of this study is to calculate the vesical central pressure, the length of the urethra is not important. The model is axisymmetric; so the bladder outlet does not have a cruciate form, but boundary conditions are applied in a way that accurately models bladder outlet during urination.

In the physiological model, urethra passes through a support structure and this structure supports the bladder. The support structure is modeled like a bowl-shape based on its anatomy [13, 14]. Janda et al. reported that pelvic diaphragm supporting bladder is an oval-shaped with 140mm of log axis dimension and 122 mm of short axis dimension [15]. To further simplify calculations, a circular bar was utilized with the radius taking the average of the 2 mentioned values, 131mm and with 2mm of thickness similar to levator ani's thickness [16].

The Eulerian kinematic description is proposed for fluid flow. A large enough fluid region is considered which encompasses all Lagrangian details. The bladder is filled with urine. Since each case is asked to sneeze deliberately during urination, the bladder is not completely full at the time of sneezing. Figure 3 shows the final computational model of the physiological bladder.

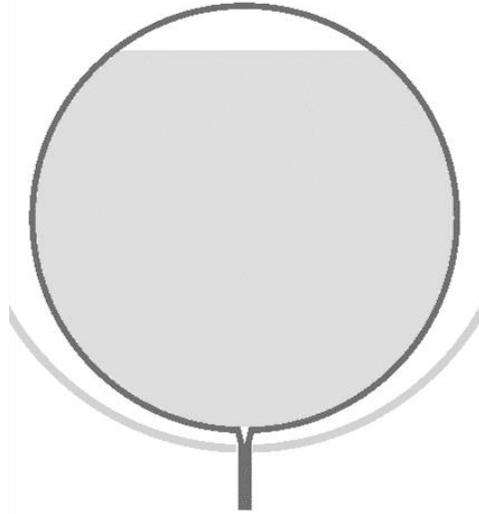

**Fig. 3** Final computational model for physiological condition

The abdominal pressure in urodynamic measurements resulting from sneezing is the most important boundary condition which is applied to the top hemisphere of the bladder. It is better to apply pressure gradually. For the first 10 seconds, the pressure remains zero. Then in the following 200 seconds, the pressure reaches its maximum level and decreases to zero level again. Rising and falling in pressure level is to model sneezing. The maximum pressure is derived from urodynamic measurements. In both pathological and physiological conditions, the abdominal pressure during deliberate sneezing is 50 cm $H_2O$ (4.9 kPa).

In order to discretize the model, a structured mesh is employed. The Lagrangian mesh including bladder, urethra and support structure (exclusively for the physiological model) is unstructured quadrilateral, while thanks to sweep method, they are homogeneously dispersed in this region. The Eulerian mesh description includes 568 elements and the Lagrangian mesh description contains 32400 elements.

In continuum mechanics, material properties dictate mechanical behavior. An interpolated function describes the field variable in each element. Usually, the relation between deformation and external forces are experimentally derived and is expressed as stress-strain equation [17, 18].

If genital and urinary tissues presumed linearly elastic, their elastic modulus can be extracted from Table 1 [19].

Table 1. Elastic properties of genital and lower urinary tract tissues (Poisson ratio = 0.45) [19]

| Tissue | Elastic modulus [MPa] | Density [kg/m$^3$] |
|---|---|---|
| Bladder | 0.05 | 1030 |
| Urethra | 3 | 1030 |
| Uterus | 0.05 | 1030 |
| Vagina | 0.005 | 1030 |
| Rectum | 0.1 | 1030 |
| Intestine | 0.1 | 1030 |
| Muscle | 2.4 | 1040 |
| Pelvic Floor | 1.2 | 1030 |
| Ligament | 1.2 | 1030 |

The current study presumed the bladder, the urethra and the support structure as linearly elastic materials. Table 1 reports the constants of the linear equations.

The urine flow was modeled linearly at first, but there was no accordance with the clinical results. So, a polynomial equation was employed to describe the urine flow. Urine density is 1g/cm$^3$ [8]. In linear modeling with neglecting thermal effects, the only incorporated parameter is shear stress modulus is 6×10$^6$ kPa for urine.

In the representative polynomial equation of fluid flow, the pressure is derived from equation (1):

Eq. (1) $W = A_1\mu + A_2\mu^2 + A_3\mu^3 + (B_0 + B_1\mu)P_0 e$

μ stands for compression, ρ$_0$ is density at P$_0$, $e$ internal force per mass unit and other constants are water constants which are presented in Table (2).

Table 2. Urine polynomial model constants [8]

| Parameter | Value |
|---|---|
| $A_1$ | $2.2 \times 10^6$ |
| $A_2$ | $9.54 \times 10^6$ |
| $A_3$ | $1.45 \times 10^6$ |
| $B_0$ | 0.28 |
| $B_1$ | 0.28 |

In order to model SUI, it is not sufficient to model only bladder and urethra. It is important to consider urine flow. Abdominal pressure increases flow pressure and as a result, urine exerts pressure on the walls of the bladder and the urethra. Fluid-structure Interactions is indispensable in computations. The shear stress at the interface leads to solid deformation or dislocation which impacts fluid flow [20-22]. To analyze this condition, the Arbitrary-Lagrangian-Eulerian equation is used, since it is simple and accurate. In this method, Euler equation is employed for fixed boundaries (fluids), Lagrange equation for moving boundaries (solids) and the Euler-Lagrange equation for the interactions [23]. Here, since sneezing is a quick dynamic incident, explicit method is employed.

ANSYS AUTODYN is used to analyze the model and solve its equations. Initial pressure in Euler region is zero. To solve fluid and solid equations together, FSI and Arbitrary-Lagrangian-Eulerian equation is employed. Penalty method is employed after re-meshing the deformed or dislocated region to save time. Therefore, the coupling of two Lagrangian and Eulerian methods automatically start. Self-interaction is activated in Lagrange region and its tolerance is 0.2. Maximum time of solving is 200 ms and the allowed tolerance of energy error is 4. Safety factor in time steps is 0.65. Strain rate in the Eulerian region is calculated based on the weight method, and Eulerian pressure is calculated based on the mean method. Also, mass scaling method is deactivated. The simulations were performed on a quad-core CPU with 6GB RAM computer.

**Results**

The comparison between pressure-time diagrams with respect to quantitative results is the most important part. Figure 4 shows the simulated pressure changes for pathological and physiological conditions.

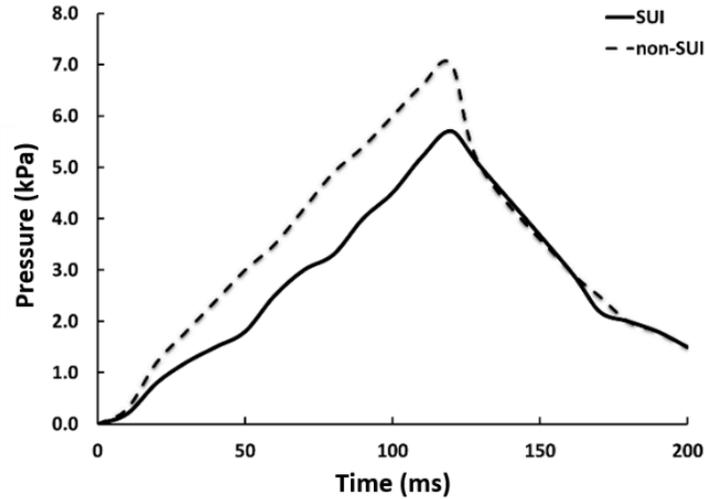

**Fig. 4** Pressure in the center of bladder in terms of time compared to physiological and pathological models

Table 3 presents the computed maximum external pressure during sneezing and urodynamic measurements and compares the results.

Table 3. Comparison of clinical bladder pressure with simulation results during a cough

| Model | Real pressure (Pa) | Simulated pressure (MPa) | Error percentage |
|---|---|---|---|
| Physiological | 6962 | 7070 | 1.3 |
| Pathological | 5785 | 5712 | 1.2 |

Bladder wall dislocation contour is drawn to study vesical deformations and is compared to clinical data. Figure 7 and 8 represents bladder wall deformation contour for the physiological and pathological model, respectively.

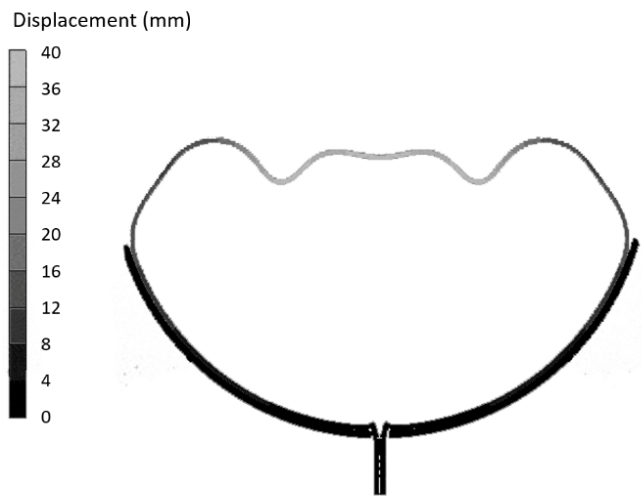

**Fig. 5** 2D Postprocessing of deformation contour in physiological model simulation (displacements in mm)

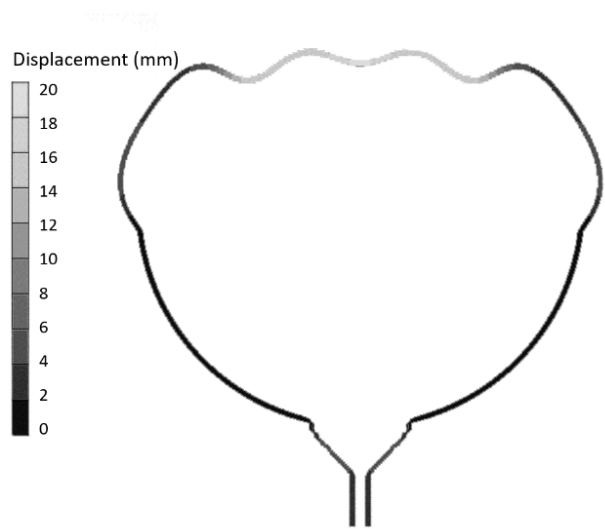

**Fig. 6** 2D Postprocessing of deformation contour in pathological model simulation (displacements in mm)

**Discussion**

Although several studies have investigated bladder and urethra mechanical behavior, their results focused on bladder filling or emptying [10, 12, 24-32]. Few studies focused on external forces and their impact on bladder condition. Zhang et al. published the most reliable results in the field of stress impact on the bladder [6]. That study modeled the involved organs during the jumping of height and presumed urine leakage occur as the urine enters the urethra. In their reported results, only the upper part of the urethra was filled with urine, while in the current study in the pathological condition, the support structure is missed and the urine fills the urethra completely which indicates that this criterion is not an appropriate criterion for urine leakage, even though urine leakage is not the preferred parameter in the present study. Similar to Zhang model [6], the current developed model did not simulate urine transmission through the urethra. Limitations in hardware resources and time-consuming processes lead to these defects. Although Zhang et al. studied 2 different stress incidents, our qualitative images of modeled bladder deformation match perfectly with Zhang et al. results [6]. Images show that elements of the upper hemisphere are deformed and dislocated more than other elements. The least dislocation occurs in the inferior parts.

Kim et al. investigated the cough impact on a 2-dimensional model of the bladder and studied pressure transmission theory. The main features of the current study are analytical methods and computational techniques which differentiate it from other studies. On the other words, abdominal pressure is modeled as a dynamic impact inserted on the superior part of bladder and bladder central pressure is calculated in the model. Kim added up abdominal pressure to bladder central pressure and bladder inner walls [5]. Kim also considered linear elastic properties. Kim's validation criterion is bladder and support structure deformation. The abdominal pressure in Kim's study is relatively equaled to clinically measured pressure for the pathological conditions in the current study. Maximum deformation of the current study is reported 2cm (Figure 6) which accords with Kim results [5].

Respecting to the diagram shown in Figure 4, bladder internal pressure in the physiological condition is more than the pathological one, which matches perfectly with urodynamic measurements. The patient diagnosed with SUI has no control on urethral pressure, so any external stress to bladder leads to urine leakage and bladder central pressure reduction. While in physiological conditions, the pressure inside the bladder increases, since support structure prevents urine leakage.

Table 3 verified quantitative simulations in the current study. It determines that the model could be verified through comparison of simulated and clinically measured data. Noise in clinical data because of environmental conditions and sensor accuracy is inevitable but simulation results are free from any noise.

For further studies, it is proposed to apply more recent FSI techniques and to define variable meshing in the Eulerian region to increase accuracy. Boundary conditions should be defined in a way that does not limit the natural activity of bladder and is more similar to the real situation. It is also proposed to utilize MRI images for geometry construction to improve the reliability of the results, in order to predict and differentiate pathological and physiological conditions.

**Conclusion**

The current study developed two models of stress urinary incontinence for physiological and pathological conditions to study urine dynamic flow in bladder under abdominal pressure resulting from sneezing. Developed model and its validations show analytical methods and simplifying biological systems, like inferior urine tract, would successfully simulate the clinical results in a virtual environment. It shows that the investigations of biological systems do not necessarily require experimental sets. Although FSI method deploys complex and time-consuming calculations, it is very practical and accurate in modeling coupled systems including both solid and fluid phase. Results indicate maximum deformation in bladder occurs when the internal pressure is reaching its maximum level. For further studies, it is proposed to utilize more recent FSI techniques, more accurate boundary conditions and geometry construction based on MRI images, in order to predict improved results in pathological and physiological conditions.

**References**

[1]  L. Wilson, J. S. Brown, G. P. Shin, K.-O. Luc, and L. L. Subak, "Annual Direct Cost of Urinary Incontinence," *Obstetrics & Gynecology,* vol. 98, pp. 398-406, 2001/09 2001.
[2]  G. Enhorning, E. R. Miller, and F. Hinman, "Simultaneous recording of intravesical and intraurethral pressure," *Acta Chirurgica candinavica,* vol. 276, pp. 1-68, 1961.
[3]  P. E. P. Petros and U. I. Ulmsten, "AN INTEGRAL THEORY OF FEMALE URINARY INCONTINENCE," *Acta Obstetricia et Gynecologica Scandinavica,* vol. 69, pp. 7-31, 1990/01 1990.
[4]  J. A. Ashton-Miller and J. O. L. Delancey, "Functional Anatomy of the Female Pelvic Floor," *Annals of the New York Academy of Sciences,* vol. 1101, pp. 266-296, 2007/02/15 2007.
[5]  K. J. Kim, "Biomechanical Analyses of Female Stress Urinary Incontinence," Ph.D., University of Michigan, 1994.
[6]  Y. Zhang, S. Kim, A. G. Erdman, K. P. Roberts, and G. W. Timm, "Feasibility of Using a Computer Modeling Approach to Study SUI Induced by Landing a Jump," *Annals of Biomedical Engineering,* vol. 37, pp. 1425-1433, 2009/05/05 2009.
[7]  B. Haridas, H. Hong, R. Minoguchi, S. Owens, and T. Osborn, "PelvicSim - A computational experimental system for biomechanical evaluation of female pelvic floor organ disorders and associated minimally invasive interventions," *Studies in Health Technology and Informatics,* vol. 119, pp. 182-187, 2006.
[8]  T. Spirka, K. Kenton, L. Brubaker, and M. S. Damaser, "Effect of Material Properties on Predicted Vesical Pressure During a Cough in a Simplified Computational Model of the Bladder and Urethra," *Annals of Biomedical Engineering,* vol. 41, pp. 185-194, 2012/08/21 2012.

[9]     L. Chan, S. The, J. Titus, and V. Tse, "P14.02: The value of bladder wall thickness measurement in the assessment of overactive bladder syndrome," *Ultrasound in Obstetrics and Gynecology,* vol. 26, pp. 460-460, 2005/09/08 2005.
[10]    M. S. Damaser and S. L. Lehman, "The effect of urinary bladder shape on its mechanics during filling," *Journal of Biomechanics,* vol. 28, pp. 725-732, 1995/06 1995.
[11]    K. A. Backman, "Effective urethral diameter," in *Hydrodynamics of Micturition*, C. C. Thomas, Ed., ed IL: Springfield, 1971, pp. 250-256.
[12]    R. A. Hosein and D. J. Griffiths, "Computer simulation of the neural control of bladder and urethra," *Neurourology and Urodynamics,* vol. 9, pp. 601-618, 1990.
[13]    R. L. Drake, W. Vogl, and A. W. Mitchell, *Gray's Anatomy for Students*. New York: Elsevier, 2005.
[14]    F. H. Netter, *Atlas of Human Anatomy*. Philadelphia: Saunders Elsevier, 2006.
[15]    Š. Janda, F. C. T. van der Helm, and S. B. de Blok, "Measuring morphological parameters of the pelvic floor for finite element modelling purposes," *Journal of Biomechanics,* vol. 36, pp. 749-757, 2003/06 2003.
[16]    D. d'Aulignac, J. A. C. Martins, E. B. Pires, T. Mascarenhas, and R. M. N. Jorge, "A shell finite element model of the pelvic floor muscles," *Computer Methods in Biomechanics and Biomedical Engineering,* vol. 8, pp. 339-347, 2005/10 2005.
[17]    S. S. Rao, "Overview of Finite Element Method," in *The Finite Element Method in Engineering*, ed: Elsevier, 2005, pp. 3-49.
[18]    H. Saunders, "Book Reviews : THE FINITE ELEMENT METHOD FOR ENGINEERS K. E. Huebner John Wiley & Sons - New York (1975)," *The Shock and Vibration Digest,* vol. 9, pp. 39-39, 1977/02/01 1977.
[19]    H. Yamada, *Strength of Biological Materials*. Baltimore: Williams and Wilkins, 1970.
[20]    E. Kuhl, S. Hulshoff, and R. de Borst, "An arbitrary Lagrangian Eulerian finite-element approach for fluid-structure interaction phenomena," *International Journal for Numerical Methods in Engineering,* vol. 57, pp. 117-142, 2003.
[21]    E. J. Weinberg and M. R. Kaazempur Mofrad, "Transient, Three-dimensional, Multiscale Simulations of the Human Aortic Valve," *Cardiovascular Engineering,* vol. 7, pp. 140-155, 2007/11/16 2007.
[22]    M. Souli, A. Ouahsine, and L. Lewin, "ALE formulation for fluid–structure interaction problems," *Computer Methods in Applied Mechanics and Engineering,* vol. 190, pp. 659-675, 2000/11 2000.
[23]    J. Donea, A. Huerta, J. P. Ponthot, and A. Rodrï¿½ï¿½guez-Ferran, "Arbitrary Lagrangian-Eulerian Methods," in *Encyclopedia of Computational Mechanics*, ed: John Wiley & Sons, Ltd, 2004.
[24]    D. J. Griffiths, "Urethral elasticity and micturition hydrodynamics in females," *Medical & Biological Engineering,* vol. 7, pp. 201-215, 1969/03 1969.
[25]    A. Spángberg, H. Terió, A. Engberg, and P. Ask, "Quantification of urethral function based on Griffiths' model of flow through elastic tubes," *Neurourology and Urodynamics,* vol. 8, pp. 29-52, 1989.
[26]    E. H. C. Bastiaanssen, J. L. van Leeuwen, J. Vanderschoot, and P. A. Redert, "A Myocybernetic Model of the Lower Urinary Tract," *Journal of Theoretical Biology,* vol. 178, pp. 113-133, 1996/01 1996.
[27]    E. H. C. Bastiaanssen, J. Vanderschoot, and J. L. van Leeuwen, "State-space Analysis of a Myocybernetic Model of the Lower Urinary Tract," *Journal of Theoretical Biology,* vol. 180, pp. 215-227, 1996/06 1996.
[28]    D. J. Griffiths, "Hydrodynamics of male micturition—I Theory of steady flow through elastic-walled tubes," *Medical & Biological Engineering,* vol. 9, pp. 581-588, 1971/11 1971.
[29]    M. Horák and J. Křen, "Mathematical model of the male urinary tract," *Mathematics and Computers in Simulation,* vol. 61, pp. 573-581, 2003/01 2003.

[30]  R. van Mastrigt and D. J. Griffiths, "An evaluation of contractility parameters determined from isometric contractions and micturition studies," *Urological Research,* vol. 14, pp. 45-52, 1986/02 1986.
[31]  J. R. Fielding, D. J. Griffiths, E. Versi, R. V. Mulkern, M. L. Lee, and F. A. Jolesz, "MR imaging of pelvic floor continence mechanisms in the supine and sitting positions," *American Journal of Roentgenology,* vol. 171, pp. 1607-1610, 1998/12 1998.
[32]  H. T. Lotz, P. Remeijer, M. van Herk, J. V. Lebesque, J. A. de Bois, L. J. Zijp*, et al.*, "A model to predict bladder shapes from changes in bladder and rectal filling," *Medical Physics,* vol. 31, pp. 1415-1423, 2004/05/24 2004.